\documentclass[
preprint,
prd,
showpacs, 
amsmath, 
amssymb, 
]{revtex4}
\usepackage[dvips]{graphicx}

\begin{document}

\title{Close-limit analysis for
head-on collision of two black holes in higher dimensions:
Brill-Lindquist initial data}

\author{Hirotaka Yoshino$^{(1)}$}
\email{yoshino@th.phys.titech.ac.jp}
\author{Tetsuya Shiromizu$^{(1,2,3)}$}
\email{shiromizu@phys.titech.ac.jp}
\author{Masaru Shibata$^{(4)}$}
\email{shibata@provence.c.u-tokyo.ac.jp}


\affiliation{$^{(1)}$ Department of Physics, Tokyo Institute of Technology, Tokyo 152-8551, Japan}

\affiliation{$^{(2)}$ Department of Physics, The University of Tokyo,  Tokyo 113-0033, Japan}

\affiliation{$^{(3)}$ Advanced Research Institute for Science and Engineering, 
Waseda University, Tokyo 169-8555, Japan}

\affiliation{$^{(4)}$ Graduate School of Arts and Sciences, University of Tokyo, Komaba, Meguro, Tokyo 153-8902, Japan}

\date{August 16, 2005}

\begin{abstract}
Motivated by the TeV-scale gravity scenarios, we study gravitational
radiation in the head-on collision of two black holes in higher
dimensional spacetimes using a close-limit approximation.  We prepare
time-symmetric initial data sets for two black holes (the so-called
Brill-Lindquist initial data) and numerically evolve the spacetime in 
terms of a gauge invariant formulation for the perturbation around the
higher-dimensional Schwarzschild black holes. The waveform and
radiated energy of gravitational waves emitted in the head-on
collision are clarified. Also, the complex frequencies of fundamental
quasinormal modes of higher-dimensional Schwarzschild black holes,
which have not been accurately derived so far, are determined. 
\end{abstract}

\maketitle

\section{Introduction}

Clarifying the nature of black holes in higher-dimensional spacetimes
has become an important issue since the possibility of the black-hole
production in accelerators was pointed out. If our space is the
3-brane in large \cite{ADD98} or warped \cite{RS99} extra
dimensions, the Planck energy could be of $O({\rm TeV})$ that is
accessible with the planned accelerators. If the number of dimension $D$ of
our spacetime is actually larger than 4, a black hole of very small 
mass will be produced artificially in particle experiments and the
evidence may be detected. 

The possible phenomenology of the black holes which may be produced in
accelerators was first discussed in \cite{BHUA} (see \cite{reviews}
for reviews).  During the high-energy particle collision of
sufficiently small impact parameter in a higher-dimensional spacetime,
two particles will form a distorted black hole of small
mass. Subsequently, it settles down to a stationary state after
emission of gravitational waves. The stationary black hole will soon be
evaporated by the Hawking radiation, indicating that the quantum
gravity effects will be important.  The evaporation and quantum
gravity effects \cite{greybody,quantum} have been studied for yielding 
a plausible scenario (cf. \cite{related} for related issues). On the
other hand, the analyses for formation of the black hole and for the
subsequent evolution by gravitational radiation are still in an early
stage.  These phases are described well in the context of general
relativity \cite{GR04} (see also \cite{R04}), but due to its highly
nonlinear nature, the detailed process has not been well understood.

More specifically, two issues should be clarified for these phases.
One is the condition (i.e., the impact parameter) for formation of
a black hole and the other is the fate of the formed black hole after
emission of gravitational waves, which can be used as the 
initial condition of the Hawking radiation phase. 

Extensive effort has been made in the past five years for the first
issue.  The popular method is to approximate the high energy particle
of no charge or spin by the Aichelburg-Sexl shock-wave metric
\cite{AS71}. The merit of this approximation is that superimposing two
Aichelburg-Sexl metrics, a metric of two
particles moving with speed of light can be derived for a
spacetime region in which causal connection between two particles is
absent. Although this solution can form a naked singularity at the
collision and it is not clear what happens after the collision,
it is still possible to determine the
condition for the formation of an apparent horizon for a spacelike
hypersurface of the known solution. Formation of the apparent horizon
is a sufficient condition for formation of a black hole for which the
event horizon is
outside the apparent horizon. Thus, the lower bound of the impact
parameter can be estimated. 
Such study was first done by Eardley and Giddings \cite{EG02} 
in a four-dimensional case and it was extended to $D$-dimensional cases
by Yoshino and Nambu \cite{YN03}. Recently
these studies were improved by Yoshino and Rychkov \cite{YR05}
by analyzing the apparent horizon
in a different spacelike hypersurface from the one in \cite{EG02, YN03}.

The second issue is to clarify the final state of a black hole formed
after collision of two particles, which will be a Kerr black hole in
higher dimensions, perhaps those described by Myers and Perry
\cite{MP86}. Here, the stationary Kerr black holes of no electric
charge are described by the mass and angular momentum. Namely, the
goal is to derive a formula of the final mass $M_{\rm final}$ and
angular momentum $J_{\rm final}$ as functions of the initial impact
parameter and initial energy of two particles.  For this issue, a
couple of preliminary analyses have been carried out so far (see
Ref. \cite{CBC05} for a review). 

One is the work by Yoshino and Rychkov \cite{YR05}, who constrained 
the allowed region of the mass and angular momentum by finding
the apparent horizon for spacetimes of two Aichelburg-Sexl particles
and subsequently employing the area theorem. However, the allowed region
cannot be pinpointed with this approach, implying that the analysis of
gravitational waves emitted during the collision is inevitable. 

The gravitational radiation from two particles with speed of light in
head-on collision was first computed by D'Eath and Payne \cite{DP92}
(summarized in \cite{D96}). They analyzed a spacetime of two
Aichelburg-Sexl particles in the $D=4$ case, paying attention only to
a region far from the particles and using a perturbative
theory. By this analysis, the radiation near the symmetric axis can be
calculated. Assuming the axisymmetric angular pattern of the radiation, 
they estimated the total radiated energy $E_{\rm rad}$ as 
16\% of the total energy of the system. 

Recently, Cardoso {\it et al.} \cite{CDL03} studied gravitational
radiation in the linear perturbation theory of the higher-dimensional
flat spacetime.  They found again that about 16\% of the total
energy will be emitted in the head-on collision for $D=4$, 
which is consistent with the results by D'Eath and Payne. 
They also found that the efficiency is highly suppressed for larger value of
$D$, e.g., about $0.001\%$ for $D=10$. 

The black hole perturbation theory has been also used recently.
Cardoso {\it et al.} \cite{CL02} computed gravitational waves from a
particle of energy $\mu$ with a speed close to the light speed falling
straightforwardly into a Schwarzschild black hole of mass $M \gg \mu$
for $D=4$. Berti {\it et al.} \cite{BCG04} extend their work for $D >
4$ (see \cite{BH-plus-lightlike} for further
generalizations). Extrapolating the results for $\mu \rightarrow M/2$,
they found that the radiation efficiency is $\approx 13\%$ for $D=4$
and decreases with increasing the value of $D$, e.g., $8$\% for
$D=10$.

Although these approximate studies could give an approximate value of
the radiation efficiency, it is natural to consider that the error in
the estimate is still factor of 2 or more. To derive the exact
numerical value, it is necessary to carry out more strict
analysis. One promising approach is to employ numerical simulation in
full general relativity. In the $D=4$ case, simulation for black hole
collision is feasible \cite{numerical_relativity}, producing certain
scientific results. However, these works have been done for the case
that velocity of each black hole is much smaller than the speed of
light. Formulation and numerical technique for black-hole collision
with a very large Lorentz factor $\gamma \gg 1$ have not been 
developed yet. 

In this paper, we adopt the so-called close-limit method for computing
gravitational radiation, which was originally developed by Price and
Pullin \cite{PP94}. In this method, we prepare two black holes of a
small separation as the initial condition. If the separation is small
enough to form a common horizon, the spacetime can be
well approximated by a perturbed black-hole spacetime. As a result,
the gravitational radiation during the collision can be analyzed in
the context of the black-hole perturbation theory. This method has
been applied for two black holes initially at rest \cite{AP96, AP97},
initially approaching with linear momentum \cite{moving}, and many
other two-black-hole systems \cite{other_closelimit}. The robustness
of this method is established by confirming that the results by this
method agree with those in numerical relativity. This fact motivates
us to adopt the close-limit approximation for high-velocity collision
of two black holes in higher dimensional spacetimes.

As a first step toward the series of study for more plausible cases,
in this paper, we focus on head-on collision with time-symmetric
initial data of two equal-mass black holes. For simplicity, we choose
the Brill and Lindquist initial data \cite{BL63} that describes a
spacelike hypersurface in a spacetime composed of three sheets
connected by two Einstein-Rosen bridges.  We will show that the
successful numerical results and indicate that extension of the
analyses with more general initial data is straightforward.

This paper is organized as follows. In the Sec. II, we introduce the
Brill-Lindquist initial data and analyze the apparent horizons for $D
\geq 4$. In Sec. III, we derive the close-limit form of the initial
data and briefly review the master equation for the perturbation of
the higher-dimensional Schwarzschild black hole \cite{KI03}. We also
explain our numerical methods.  Numerical results are shown in
Sec. IV, paying attention to the radiated energy and gravitational
waveforms. Section V is devoted to a summary.  In appendix A, the
gauge-invariant perturbation formalism as well as a method for
preparing initial master variable from the Brill-Lindquist initial
data are presented. Appendix B describes a formula for computing the
radiated energy of gravitational waves from the master variable.

\section{The Brill-Lindquist initial data}

\subsection{The Brill-Lindquist two-black-hole solution}

Let $(\Sigma, h_{ab},K_{ab})$ denote a $(D-1)$-dimensional
spacelike hypersurface $\Sigma$ with the metric $h_{ab}$
and the extrinsic curvature $K_{ab}$ in a $D$-dimensional spacetime.
The equations of the Hamiltonian and momentum constraints are
\begin{equation}
{}^{(n+1)}R-h^{ab}h^{cd}K_{ac}K_{bd}+K^2=0, 
\end{equation}
\begin{equation}
\nabla^a(K_{ab}-h_{ab}K)=0,
\end{equation}
where ${}^{(n+1)}R$ is the Ricci scalar of $\Sigma$, 
$\nabla^a$ is the covariant derivative with respect to $h_{ab}$, and
$n=D-2$.
Assuming the time symmetry (i.e., $K_{ab}=0$), Eq. (2)
is satisfied trivially. Assuming further the conformal flatness
$h_{ab}=\varPsi^{4/(n-1)}\delta_{ab}$, 
the Hamiltonian constraint equation is written to the Laplace equation
for the conformal factor
\begin{equation}
\nabla_{\rm f}^2\varPsi=0,
\label{eq:Laplace}
\end{equation}
where $\nabla_{\rm f}^2$ is the flat-space Laplacian. 

We introduce the cylindrical coordinates $(\rho, z)$ in which
the flat-space metric is given by 
$ds_{\rm f}^2=dz^2+d\rho^2+\rho^2d\Omega_{n-1}^2$
with the metric $d \Omega_{n-1}^2$ on the $(n-1)$-dimensional unit sphere. 
Among an infinite number of solutions for Eq.~\eqref{eq:Laplace} that
denote spacetimes of two black holes, we choose the following one
composed of two point sources located at $z=\pm z_0$ along the $z$-axis as 
\begin{equation}
\varPsi=1+\frac18{[r_h(M)]^{n-1}}
\left(\frac{1}{R_-^{n-1}}+\frac{1}{R_+^{n-1}}\right),
\label{BL_solution}
\end{equation}
where $R_{\pm}\equiv \sqrt{(z\mp z_0)^2+\rho^2}$,
$M$ is the total gravitational mass of the system, 
and $r_h(M)$ is the gravitational radius defined by
\begin{equation}
r_h(M)=\left(\frac{16\pi GM}{n\Omega_n}\right)^{1/(n-1)}. 
\end{equation}  
Here, $\Omega_n=2\pi^{(n+1)/2}/\Gamma((n+1)/2)$ is the $n$-dimensional
area of a unit sphere. Hereafter, we adopt $r_h(M)$ as the unit of the
length. The solution \eqref{BL_solution} provides the system of three
sheets connected by two Einstein-Rosen bridges.  $r_{\pm}=0$ and $r 
\rightarrow \infty$ correspond to spatial infinities of each sheet as 
found by Brill and Lindquist~\cite{BL63} for $D=4$.

\subsection{Analysis for apparent horizon}

The close-limit approximation holds for the system sufficiently close
to a stationary one-black-hole spacetime. Thus, this method can be
applied only for the case that a common apparent horizon surrounding
two black holes is present. Because of this reason, it is necessary to
clarify the range of $z_0$ for which a common apparent horizon exists. 
Also, by obtaining the area of the common apparent horizon, we can
estimate the lower bound of the final mass 
as well as the upper bound of the energy radiated away by
gravitational waves, using the area theorem of black holes. 

The common apparent horizon is determined by a numerical method
developed by Yoshino and Nambu \cite{YN04}. Since the system is
axisymmetric, it is easily determined by a simple shooting method. The
shape of the common apparent horizon changes from a sphere at $z_0=0$ 
to a spheroid for $z_0>0$, increasing the ellipticity. 
At a critical value, $z_0^{\rm (crit)}$, it disappears.
The values of $z_0^{\rm (crit)}$ are summarized in Table~\ref{z0-crit}.

\begin{table}[tb]
\caption{The critical values of $z_0^{\rm (crit)}$ for formation of 
a common apparent horizon for $4\leq D \leq 11$.  }
\begin{ruledtabular}
\begin{tabular}{c|cccccccc}
$D$ & $4$ & $5$ & $6$ & $7$ & $8$ & $9$ & $10$ & $11$  \\
  \hline 
$z_0^{\rm (crit)}/r_h(M)$ & $0.192$ & $0.393$ & $0.510$ & $0.586$ & $0.641$ & $0.683$ & $0.715$ & $0.742$ 
\end{tabular}
  \end{ruledtabular}
  \label{z0-crit}
\end{table}

\begin{figure}[tb]
\centering
{
\includegraphics[width=0.5\textwidth]{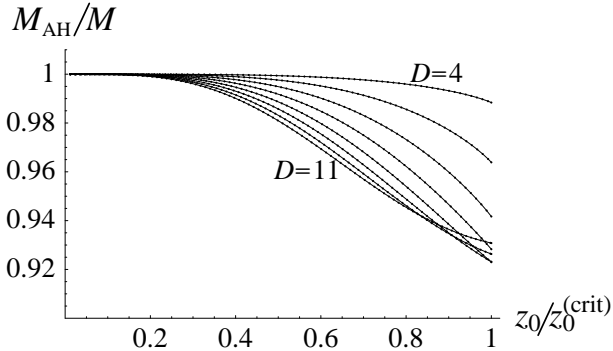}}
\caption{The relation between $z_0/z_0^{\rm (crit)}$ and the trapped
energy $M_{\rm AH}/M$ at the initial state for $D=4$--11. The plot
range of the vertical axis is $0.9\le M_{\rm AH}/M\le 1$.}
\label{MAHD_initial}
\end{figure}

In the presence of the common apparent horizon, 
the mass of the apparent horizon is defined by
\begin{equation}
M_{\rm AH}=\frac{n\Omega_n}{16\pi G}
\left(\frac{A_{\rm AH}}{\Omega_n}\right)^{(n-1)/n},
\label{AHmass-def}
\end{equation}
where $A_{\rm AH}$ denotes the $n$-dimensional area of the apparent
horizon. In the Brill-Lindquist initial data, $M_{\rm AH}$ coincides
with the Hawking quasilocal mass \cite{SH68} evaluated on the horizon
and indicates the trapped energy at the initial state. $M_{\rm AH}$
provides us the lower bound of the final mass $M_{\rm final}=M-E_{\rm
rad}$ where $E_{\rm rad}$ is the radiated energy of gravitational
waves. Equivalently $M-M_{\rm AH}$ gives the upper bound of $E_{\rm
rad}$. Figure~\ref{MAHD_initial} shows the relation between
$z_0/z_0^{\rm (crit)}$ and $M_{\rm AH}/M$. In the $D=4$ case, about
99\% of the total energy is trapped inside the apparent horizon at the
initial state. On the other hand, the trapped energy $M_{\rm AH}/M$ 
becomes smaller for larger values of $D$. 

\section{The close-limit analysis}

\subsection{The close-limit of the Brill-Lindquist initial data}

In this paper, the spacetime of two black holes is evolved using the
close-limit approximation, in which the evolution of the field
variables are carried out by a gauge-invariant perturbation technique. 
For this analysis, it is necessary to derive an initial condition for
the linear theory in the Schwarzschild background. To do so, the 
Brill-Lindquist metric is rewritten to 
\begin{equation}
ds^2=\varPsi^{4/(n-1)}[dR^2+R^2(d\theta^2+\sin^2\theta d\Omega_{n-1}^2)],
\end{equation}
\begin{equation}
\varPsi=1+\frac{1}{4R^{n-1}}+\frac{1}{4R^{n-1}}\sum_{l=2,4,\cdots}
\left(\frac{z_0}{R}\right)^l
C_l^{[(n-1)/2]}(\cos\theta),
\end{equation}
where $R=\sqrt{z^2+\rho^2}$, $\theta=\tan^{-1}(\rho/z)$, and
$C_l^{[\lambda]}$ denotes the Gegenbauer polynomials which
are defined by the generating function 
\begin{equation}
\left(1-2xt+t^2\right)^{-\lambda}=\sum_{l=0}^{\infty}C_l^{[\lambda]}(x)t^l.
\end{equation}
Here, we assume $z_0 \ll r_h(M)=1$. 
If $z_0=0$, the metric provides the space component of
the Schwarzschild metric in the isotropic coordinate. 

We introduce a new coordinate $r$ through 
the ordinary relation between the Schwarzschild coordinate $r$ 
and the isotropic coordinate $R$:
\begin{equation}
r=R\varPsi_0^{2/(n-1)}, ~~~~\varPsi_0=1+{1}/{4R^{n-1}},
\end{equation}
or equivalently
\begin{equation}
R=\left[\left(r^{(n-1)/2}+\sqrt{r^{n-1}-1}\right)/2\right]^{2/(n-1)}.
\label{isotropic}
\end{equation}
Then the metric becomes 
\begin{equation}
ds^2=\left(\frac{\varPsi}{\varPsi_0}\right)^{4/(n-1)}
\left[\frac{dr^2}{1-1/r^{n-1}}+r^2\left(d\theta^2+\sin^2\theta d\Omega_{n-1}^2\right)\right],
\label{BL_Schwarzschild}
\end{equation}
\begin{equation}
\frac{\varPsi}{\varPsi_0}=
1+\frac{1/4R^{n-1}}{1+1/4R^{n-1}}
\sum_{l=2,4,\cdots}\left(\frac{z_0}{R}\right)^l
C_l^{[(n-1)/2]}(\cos\theta).
\end{equation}
Here, the metric in the square brackets of
Eq. \eqref{BL_Schwarzschild} denotes the space part of the
Schwarzschild metric. The difference of $\varPsi/\varPsi_0$ from unity
is of $O(\epsilon^2)$ where $\epsilon\equiv z_0/R$. Since the region
of $r\ge 1$ corresponds to $R\ge R_h\equiv 4^{-1/(n-1)}$,
the system can be
regarded as the Schwarzschild black hole plus its perturbation
for a sufficiently small value of $z_0$ (or $\epsilon$). 

The first order perturbation includes the mode $l=2$, 4, $\cdots$ whose order
is $O(\epsilon^l)$.  We only consider the leading $O(\epsilon^2)$
correction which is the $l=2$ mode. Then we find the prefactor
$(\varPsi/\varPsi_0 )^{4/(n-1)}$ of Eq. \eqref{BL_Schwarzschild}
becomes
\begin{equation}
\left(\frac{\varPsi}{\varPsi_0}\right)^{4/(n-1)}\simeq
1+\frac{1/(n-1)R^{n-1}}{1+1/4R^{n-1}}
\left(\frac{z_0}{R}\right)^2
C_2^{[(n-1)/2]}(\cos\theta).
\label{fac_BL}
\end{equation}
This provides the major parts of the initial data for the linear
perturbation theory in the close-limit approximation.

\subsection{Master equations and initial master variables}

In order to analyze the time evolution of the perturbation,
we use a gauge invariant formulation 
of the higher-dimensional Schwarzschild perturbation~\cite{KI03}. 
The master equation for the three types
of perturbation variables, i.e., the scalar, vector, tensor
variables was derived in Ref. \cite{KI03}. Since 
the Brill-Lindquist initial data is axisymmetric with no rotation, 
we only need to evolve one master variable of the scalar mode.

In the scalar-mode perturbation, the master variable $\Phi$ that is
related to the gauge-invariant quantities obeys the master equation
\begin{equation}
\frac{\partial^2\Phi}{\partial t^2}-\frac{\partial^2\Phi}{\partial r_*^2}
+V_S\Phi=0,
\end{equation}
where
\begin{equation}
V_S(r)=\frac{f(r)Q(r)}{16r^2H^2(r)},
\end{equation}
\begin{equation}
f(r)=1-x,~~~H(r)=m+(1/2)n(n+1)x,
\end{equation}
\begin{equation}
m=k^2-n,~~~k^2=l(l+n-1),~~~x=1/r^{n-1},
\label{kk-and-x}
\end{equation}
\begin{multline}
Q(r)=n^4(n+1)^2x^3
+n(n+1)\left[4(2n^2-3n+4)m+n(n-2)(n-4)(n+1)\right]x^2\\
-12n\left[(n-4)m+n(n+1)(n-2)\right]mx+16m^3+4n(n+2)m^2.
\end{multline}
$r_*$ denotes the tortoise coordinate defined by
\begin{equation}
r_*=\int dr/f.
\end{equation}
More explicitly, 
\begin{multline}
r_*=
r-\frac{2}{n-1}\sum_{m=1}^{n/2-1}\sin\frac{2m\pi}{n-1}
\left[
\arctan\left(-\cot\frac{2m\pi}{n-1}+r\csc\frac{2m\pi}{n-1}\right)
-\pi/2
\right]\\
+\frac{1}{n-1}\left[
\log(r-1)+
\sum_{m=1}^{n/2-1}\cos\frac{2m\pi}{n-1}
\log\left(1+r^2-2r\cos\frac{2m\pi}{n-1}\right)
\right],
\end{multline}
for even $n$ and
\begin{multline}
r_*=
r-\frac{2}{n-1}\sum_{m=1}^{(n-3)/2}\sin\frac{2m\pi}{n-1}
\left[
\arctan\left(-\cot\frac{2m\pi}{n-1}+r\csc\frac{2m\pi}{n-1}\right)
-\pi/2
\right]\\
+\frac{1}{n-1}\left[
\log\left(\frac{r-1}{r+1}\right)+
\sum_{m=1}^{(n-3)/2}\cos\frac{2m\pi}{n-1}
\log\left(1+r^2-2r\cos\frac{2m\pi}{n-1}\right)
\right],
\end{multline}
for odd $n$.

\begin{figure}[tb]
\centering
{
\includegraphics[width=0.5\textwidth]{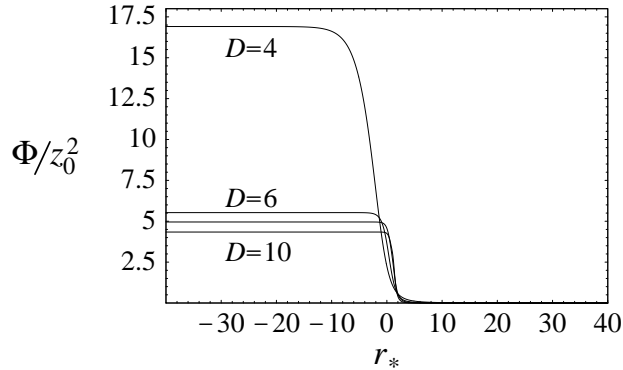}
}
\caption{The initial condition for the
master variable $\Phi/z_0^2$ as a function of $r_*$. 
The cases $D=4$, 6, 8, and 10 are shown.}
\label{initial_phi}
\end{figure}

The initial condition for the master variable $\Phi$ in gauge
invariant perturbation theory is calculated from the
metric~\eqref{BL_Schwarzschild} and \eqref{fac_BL} numerically (see
Appendix A for details). The initial values of $\Phi/z_0^2$ for
$D=4$, 6, 8, and 10 are shown in Fig.~\ref{initial_phi}. $\Phi/z_0^2$
asymptotes to a nonzero value for $r_*\to-\infty$ and to zero for
$r_*\to \infty$. It rapidly changes around $r_*\sim 0$.

\subsection{Numerical methods and numerical error}

\begin{figure}[tb]
\centering
{
\includegraphics[width=0.5\textwidth]{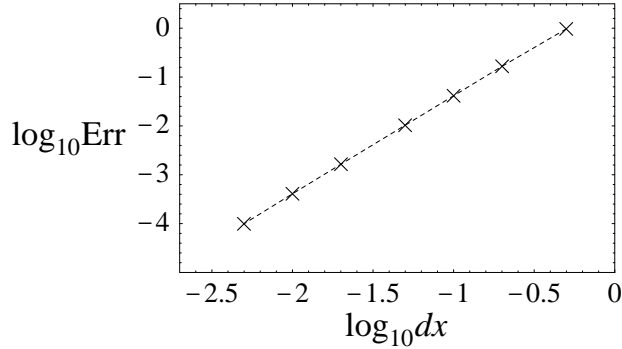}
}
\caption{The relation between the grid spacing $dx$ of $r_*$
and the numerical error. Our code shows the second-order convergence.}
\label{error}
\end{figure}

In the numerical computation for Eq. (15), we used the Lax-Wendroff
scheme which is second-order accurate in time and space.  To
validate our code, the following convergence test was carried out:
We evolved the Gaussian wave packet using the initial condition as
$h(0, r_*)=\exp(-r_*^2/100)$ with various grid resolution. Denoting
the grid spacings of $r_*$ and $t$ by $dx$ and $dt$, respectively,
there are four free parameters in our numerical code: $dx$,
$\lambda\equiv dt/dx$, and the locations of the inner and outer
boundaries $r_*^{\rm (in)}$ and $r_*^{\rm (out)}$.  First we computed
the fiducial solution $\hat{h}_{N}\equiv \Phi/z_0^2$ ($N$ denotes the
step size of the $t$ direction) at $r_*=100$ for $0\le t\le 250$ in
the $D=4$ case, choosing $dx=10^{-3}$, $\lambda=0.2$, $r_*^{\rm
(in)}=-200$, and $r_*^{\rm (out)}=200$. Then, we repeated the
computation choosing the larger values of $dx$ while fixing the other
parameters, and estimated the error by
\begin{equation}
{\rm Err}=\frac{\sum_N|h_{N}-\hat{h}_{N}|}{\sum_N|\hat{h}_{N}|}
\end{equation}
for each value of $dx$. Figure~\ref{error} shows the relation 
between $\log_{10}dx$ and $\log_{10}{\rm Err}$.
All points are located on a straight line of which slope is two. 
This illustrates the second-order accuracy of our code.

Then we analyzed the time evolution of the Brill-Lindquist initial
data by solving the master variable $\Phi/z_0^2$ starting with the
initial values shown in Fig.~\ref{initial_phi} for $4 \leq D \leq 11$.
We used $dx=0.01$ for $D=4$--$7$ and $dx=0.005$ for $D=8$--11. For the
larger values of $D$, we choose the smaller grid spacing since the
error increases with increasing $D$.  We chose the other parameters to
be $\lambda=0.2$, $r_*^{\rm (out)}=1000$, and $r_*^{\rm (in)}=-200$.
Comparing the results with those computed in poorer resolutions, we
found that the numerical error is within 0.02\% for all values of $D$.


\section{Numerical results}

\begin{figure}[tb]
\centering
{
\includegraphics[width=0.45\textwidth]{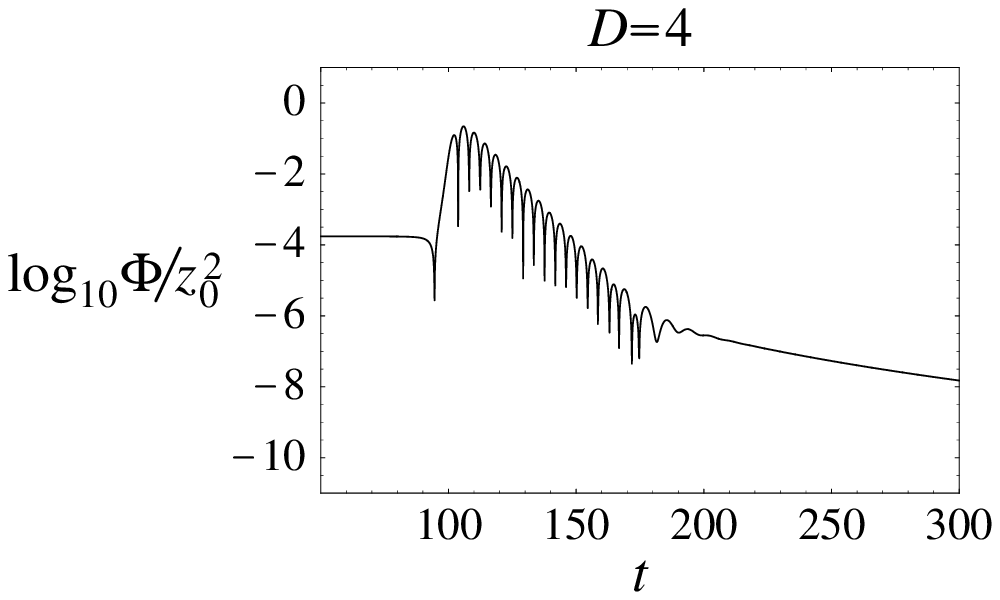}
\includegraphics[width=0.45\textwidth]{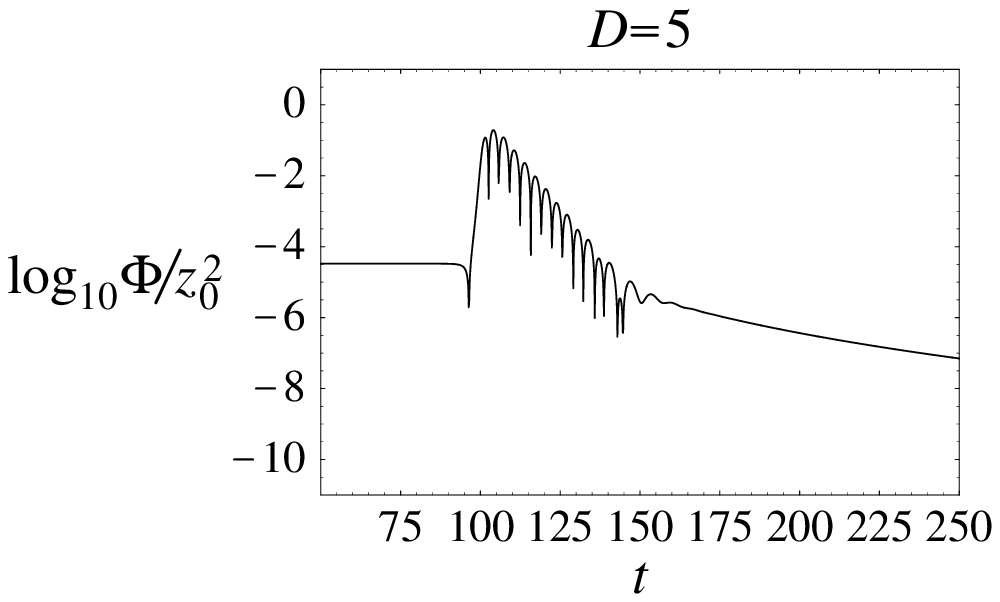}
\includegraphics[width=0.45\textwidth]{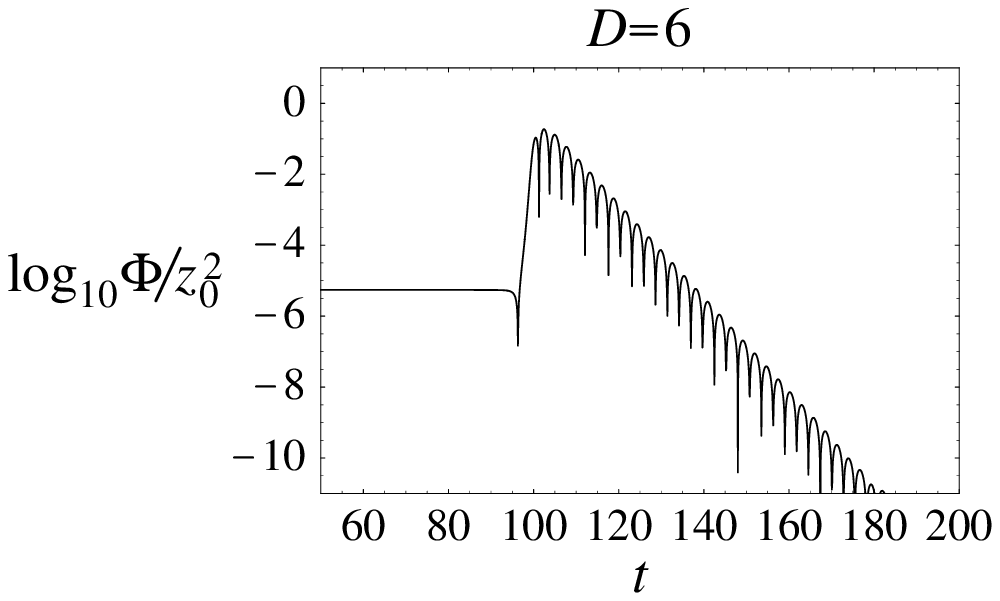}
\includegraphics[width=0.45\textwidth]{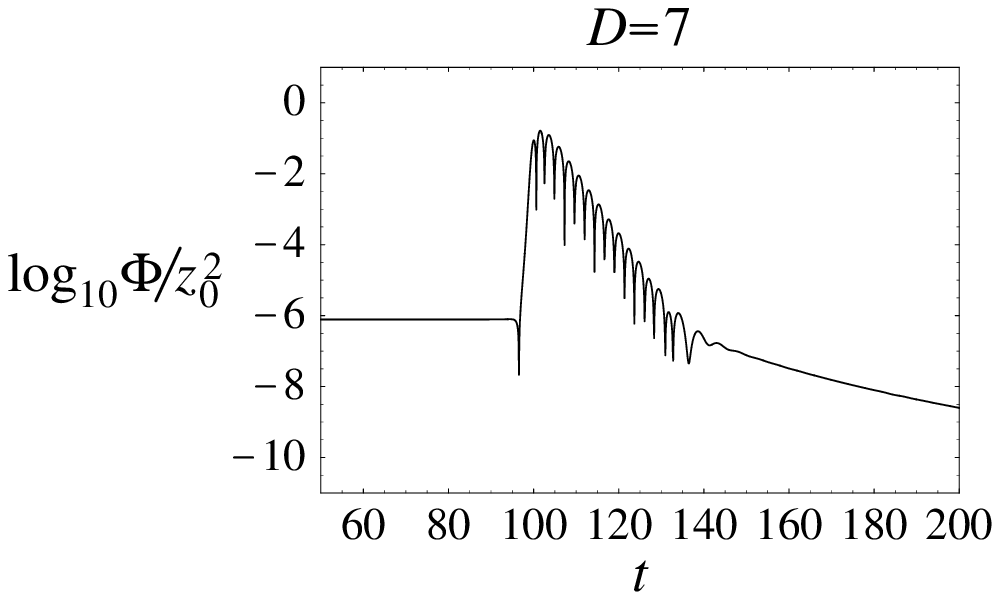}
\includegraphics[width=0.45\textwidth]{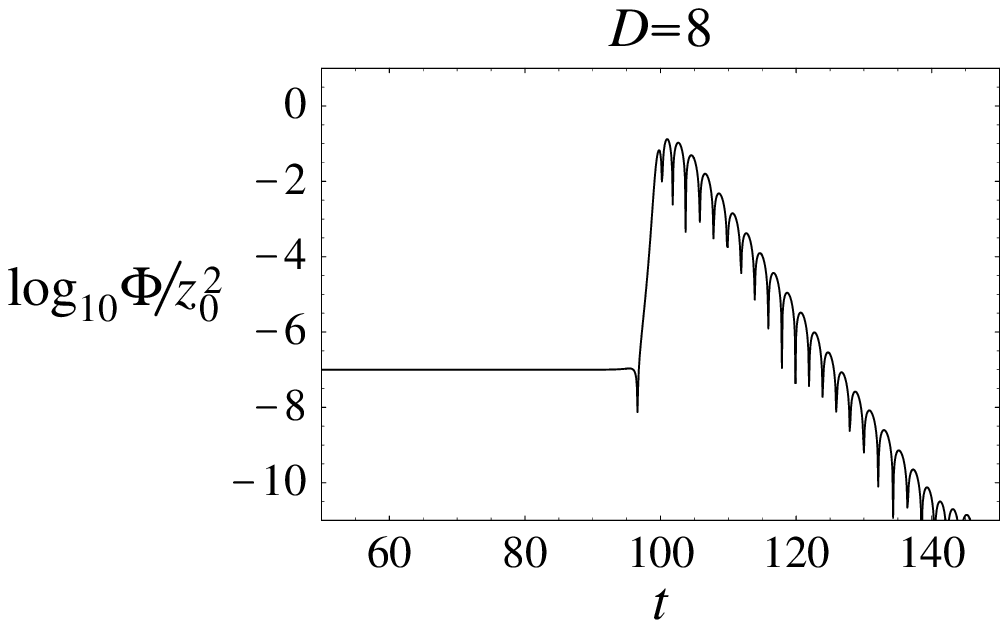}
\includegraphics[width=0.45\textwidth]{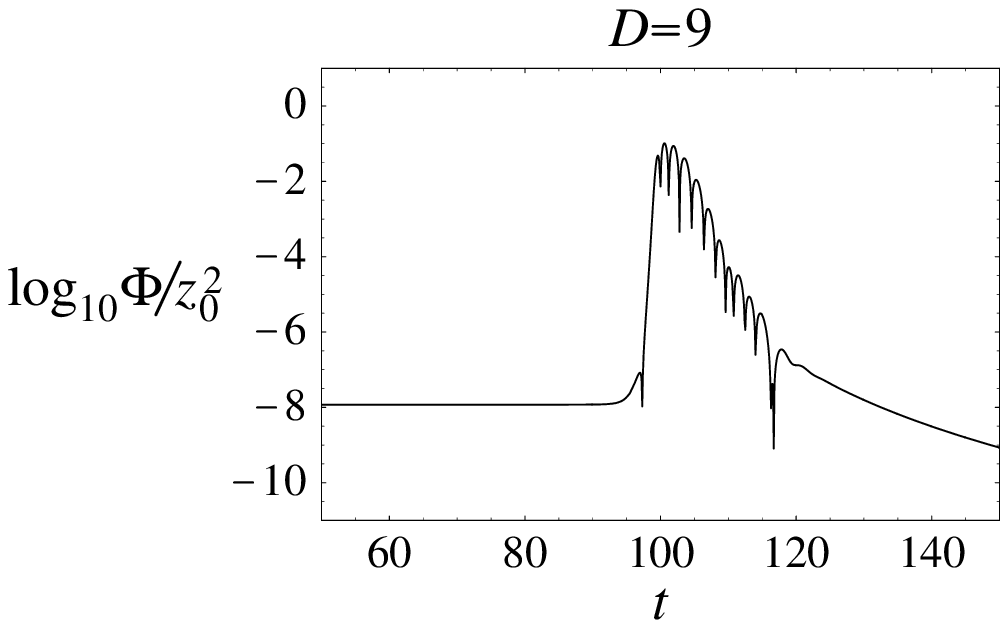}
\includegraphics[width=0.45\textwidth]{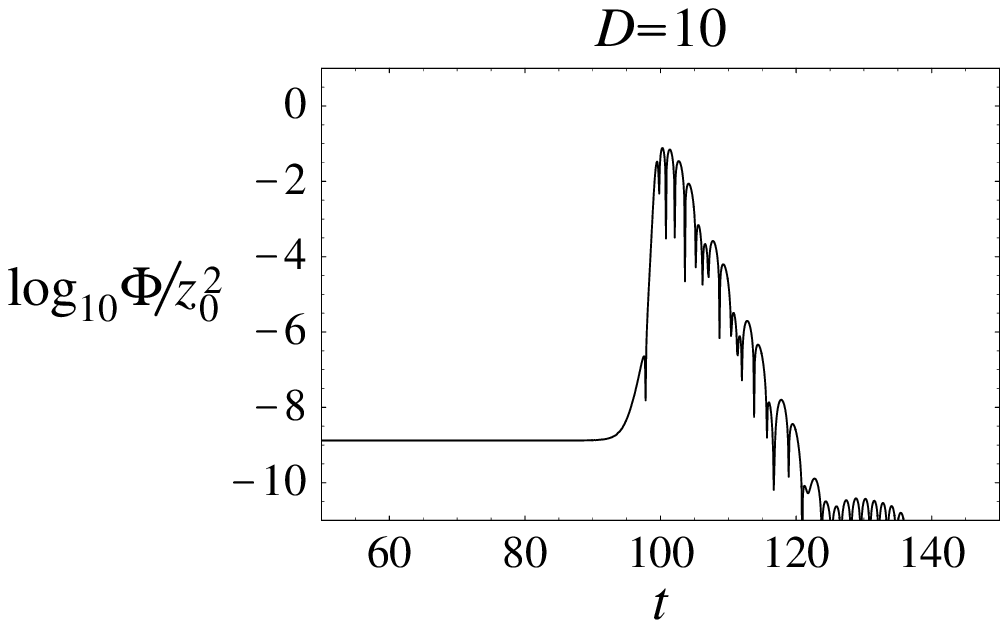}
\includegraphics[width=0.45\textwidth]{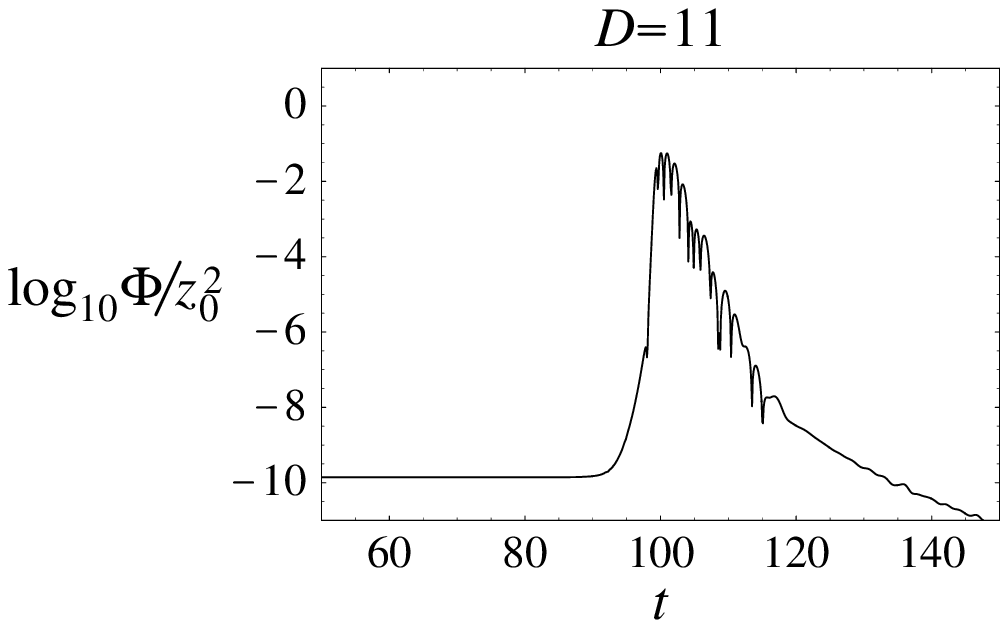}
}
\caption{Time evolution of the master variable $\Phi/z_0^2$ observed
at $r_*=100$.}
\label{evolution_master_variable}
\end{figure}

\subsection{Waveform and radiated energy}

Figure \ref{evolution_master_variable} shows the behavior of
$\Phi/z_0^2$ for $D=4$--11 observed at $r_*=100$. The quasi-normal
mode ringing is seen for all values of $D$. The power-law tail is also
computed well for $D=4$ and odd $D$. The reason that the tail cannot
be seen for even values of $D \ge 6$ is that the tail decays more
rapidly than that of odd values of $D$, as clarified in \cite{CYDL03}.

The total radiated energy $E_{\rm rad}$ is calculated by
\begin{equation}
\frac{E_{\rm rad}}{M}=
\sum_l\frac{k^2(n-1)(k^2-n)}{2n^2\Omega_n}\int\dot{\Phi}^2dt,
\label{radiated_energy_master_variable}
\end{equation}
(see Appendix B for a sketch of the derivation). 
Since $\Phi$ is proportional to $z_0^2$ for the $l=2$ mode,
$E_{\rm rad}$ is written as
\begin{equation}
{E_{\rm rad}}/{M}\simeq\hat{E}_2(z_0)^4. 
\label{radiated-E}
\end{equation}

\begin{table}[tb]
\caption{The values of $\hat{E_2}\equiv E_{\rm rad}/z_0^4$
and $E_{\rm rad}$ at $z=z_0^{\rm (crit)}$ for $D=4$--11.  }
\begin{ruledtabular}
\begin{tabular}{c|cccccccc}
$D$ & $4$ & $5$ & $6$ & $7$ & $8$ & $9$ & $10$ & $11$  \\
  \hline 
$\hat{E}_2$ & $0.0252$ & $0.0245$ & $0.0290$ & $0.0288$ & $0.0258$ & $0.0224$ & $0.0195$ & $0.0172$ \\
$E_{\rm rad}(z_0^{\rm (crit)})$ (\%) & $0.0034$ & $0.059$ & $0.20$ & $0.34$ & $0.44$ & $0.49$ & $0.51$ & $0.52$ \\ 
  \end{tabular}
  \end{ruledtabular}
  \label{Ehat}
\end{table}

Table~\ref{Ehat} shows the values of $\hat{E}_2$.  In the $D=4$ case,
$\hat{E}_2$ has been already obtained by Abrahams and
Price~\cite{AP96} as $\hat{E}_2=0.0251$. This agrees well with our numerical 
result.

To compare the radiation efficiency $E_{\rm rad}/M$ among the 
different values of $D$, one has to specify the values of $z_0$.
In Table~\ref{Ehat}, we summarize the values at 
the critical values $z_0=z_0^{\rm (crit)}$
for formation of the common apparent horizon. It is found that 
$E_{\rm rad}(z_0^{\rm (crit)})$ increases by increasing the value of $D$.
However we also should mention that 
the higher-order correction might be large for $z_0=z_0^{\rm (crit)}$. 
As we can see from Eq.~\eqref{fac_BL}, 
the characteristic value of the first order perturbation is 
$(\varPsi/\varPsi_0)^{4/(n-1)}-1$, which becomes maximal 
at the pole on the horizon. Such a maximal value is quite large, e.g., $\sim 1$ for $D=4$ and
$\sim 6$ for $D=10$. Although the close-limit method
gives a fairly good approximation beyond the regime of
the perturbation in the four-dimensional case~\cite{PP94}, further investigations
such as the second-order analysis or the full numerical
simulation are necessary to clarify this point in higher-dimensional cases.

\begin{figure}[tb]
\centering
{
\includegraphics[width=0.5\textwidth]{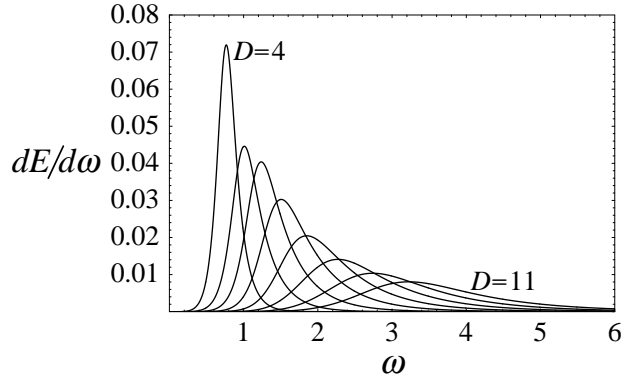}
}
\caption{The energy spectrum of gravitational waves. The unit of the vertical
axis is $Mz_0^4$. The location of the peak shifts to the
right-hand side as the value of $D$ increases.} 
\label{spectrum}
\end{figure}

Figure \ref{spectrum} shows the energy spectrum of gravitational
waves.  The value of $\omega$ at the peak becomes larger as the value
of $D$ increases.  This reflects the fact that the real part of the
fundamental quasinormal mode frequency increases with $D$ (see Sec. IV C 
for the quasinormal mode frequencies).

\begin{figure}[tb]
\centering
{
\includegraphics[width=0.5\textwidth]{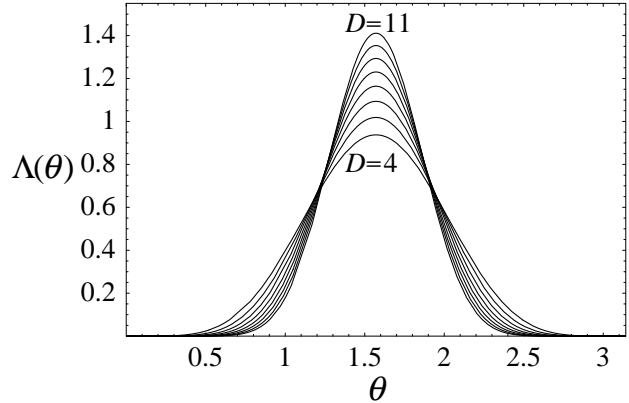}
}
\caption{The angular dependence $\Lambda(\theta)$ 
of the radiated energy. There is a peak at $\theta=\pi/2$
and $\Lambda(\pi/2)$ becomes larger as $D$ increases.}
\label{angular}
\end{figure}

The angular dependence of the radiated energy is given by
\begin{equation}
\frac{1}{E}\frac{dE}{d\theta}:=\Lambda(\theta)=
\frac{\Gamma((n+5)/2)}{\sqrt{\pi}\Gamma(2+n/2)}
\sin^{n+3}\theta,
\label{angular-dependence}
\end{equation}
(see Appendix B for a derivation).  Figure~\ref{angular} shows the
behavior of the function $\Lambda(\theta)$ for
$D=4,...,11$. Gravitational waves are mainly emitted around the
equatorial plane and this tendency is enhanced for larger $D$. This
reflects the fact that there are more directions transverse to the
symmetry axis for larger values of $D$.

\subsection{Relation between $E_{\rm rad}$ and $M_{\rm AH}$}

\begin{table}[tb]
\caption{The values of $\alpha_D$ evaluated at $z_0\ll 1$ for $D=4,...,11$.  }
\begin{ruledtabular}
\begin{tabular}{c|cccccccc}
$D$ & $4$ & $5$ & $6$ & $7$ & $8$ & $9$ & $10$ & $11$  \\
  \hline 
$\alpha_D$ & $0.0034$ & $0.016$ & $0.024$ & $0.024$ & $0.021$ & $0.017$ & $0.014$ & $0.011$ 
  \end{tabular}
  \end{ruledtabular}
  \label{alphaD}
\end{table}

\begin{figure}[tb]
\centering
{
\includegraphics[width=0.45\textwidth]{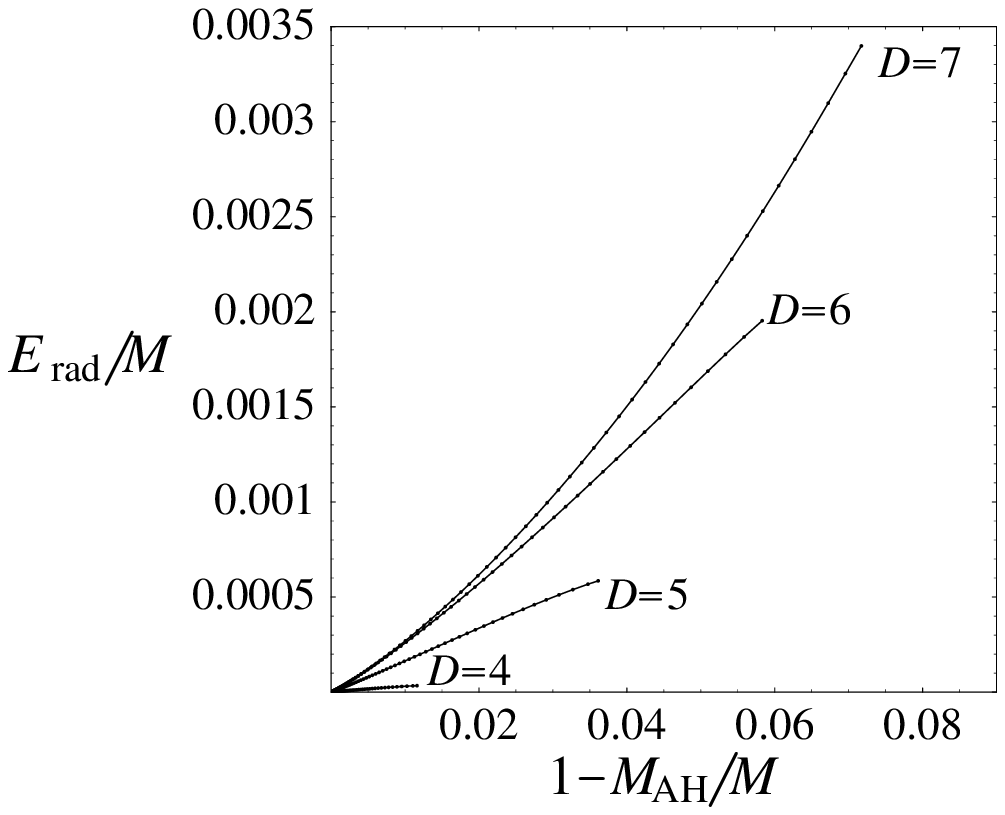}
\includegraphics[width=0.45\textwidth]{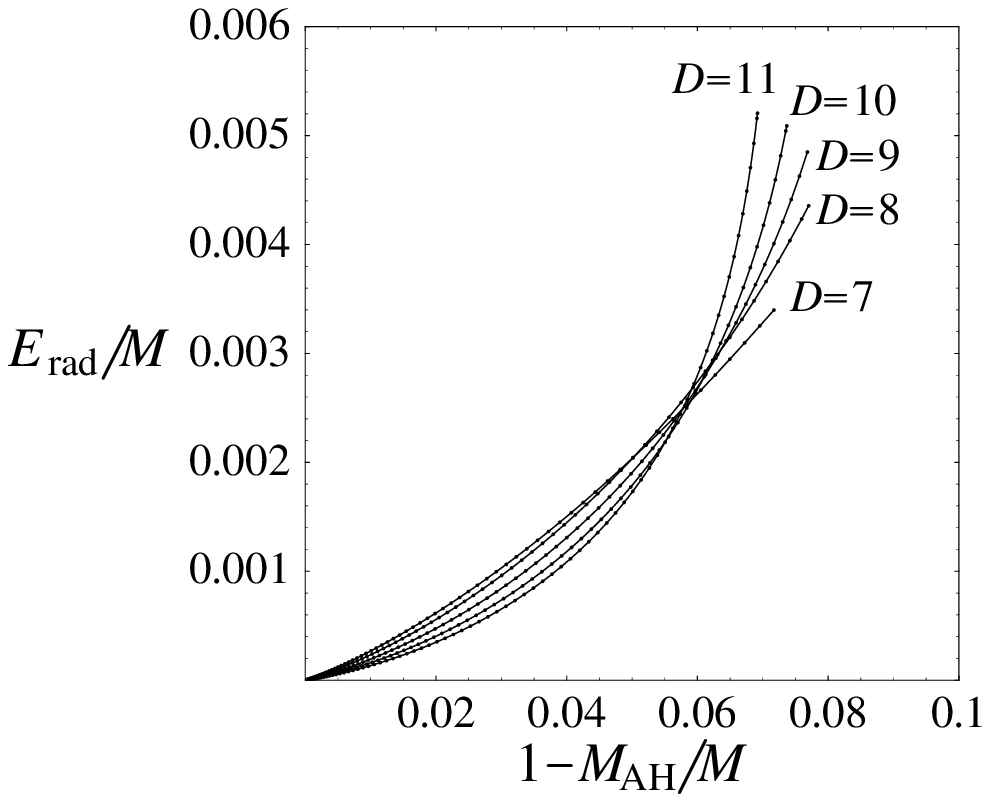}
}
\caption{The relation between $M-M_{\rm AH}$ and $E_{\rm rad}$
for $D=4,...,7$ (left) and for $D=7,...,11$ (right). The values
of $E_{\rm rad}/(M-M_{\rm AH})$ are much smaller than unity.
This is consistent with the area theorem.}
\label{AHMD_Erad}
\end{figure}

As already mentioned, $M-M_{\rm AH}$ provides the upper bound of
$E_{\rm rad}$. Hence the ratio of $E_{\rm rad}$ to $M-M_{\rm AH}$
\begin{equation}
\alpha_D\equiv\frac{E_{\rm rad}}{M-M_{\rm AH}}
\label{alpha_D}
\end{equation}
should be smaller than unity and it provides one consistency check of
our calculation. For small value of $z_0$, both $E_{\rm rad}$ and
$M-M_{\rm AH}$ are proportional to $z_0^4$ and $\alpha_D$ take nonzero
values. The values of $\alpha_D$ are summarized in Table \ref{alphaD}
and then we confirm that $\alpha_D$ is less than unity. The values of
$\alpha_D$ in higher dimensions are larger compared to $\alpha_4$ for
four dimensions.  The relation between $M-M_{\rm AH}$ and $E_{\rm
rad}$ for $0\le z_0\le z_0^{\rm (crit)}$ is also depicted in
Fig~\ref{AHMD_Erad}.  $\alpha_D\ll 1$ also holds for this parameter
range.

\subsection{Quasinormal modes}

From the ring down phase seen in Fig.~\ref{evolution_master_variable},
it is possible to derive the complex frequencies of the fundamental
quasinormal modes $\omega_{\rm QN}$.  By comparing them with previous
studies, we check the reliability of part of our results.  In
addition, the values of $\omega_{\rm QN}$ that have not been
accurately computed for large values of $D$ so far are derived from
our numerical results.

\begin{table}[tb]
\centering
\caption{The fundamental quasinormal mode of the scalar gravitational 
perturbation in the Schwarzschild black hole for $l=2$ and $4 \leq D \leq 11$. 
Results by our method, by Leaver's method, and by the WKB method are shown. }
\begin{ruledtabular}
\begin{tabular}{c|ccc}
$D$ & Our estimate  & Leaver's method & WKB   \\
  \hline 
4 & $0.747-0.177i$ & $0.7473-0.1779i$ & $0.746-0.178i$\\
5 & $0.947-0.256i$ & $0.9477-0.2561i$ & $\cdots$ \\
6 & $1.139-0.305i$ & $\cdots$ & $1.131-0.386i$\\
7 & $1.339-0.400i$ & $\cdots$ & $\cdots$ \\
8 & $1.537-0.587i$ & $\cdots$ & [$1.778-0.571i$]\\
9 & $\left\{\begin{array}{c}
1.19-0.95i\\
1.98-0.90i
\end{array}\right.$ & $\cdots$ & $\cdots$\\
10 & $\left\{\begin{array}{c}
1.25-0.94i\\
2.47-0.99i
\end{array}\right.$ & $\cdots$ & [$2.513-0.744i$] \\
11 & 
$\left\{\begin{array}{c}
1.20-0.98i\\
2.91-1.11i
\end{array}\right.$ & $\cdots$ & $\cdots$\\
\end{tabular}
\end{ruledtabular}
\label{QNM-l2}
\end{table}

\begin{figure}[tb]
\centering
{
\includegraphics[width=0.3\textwidth]{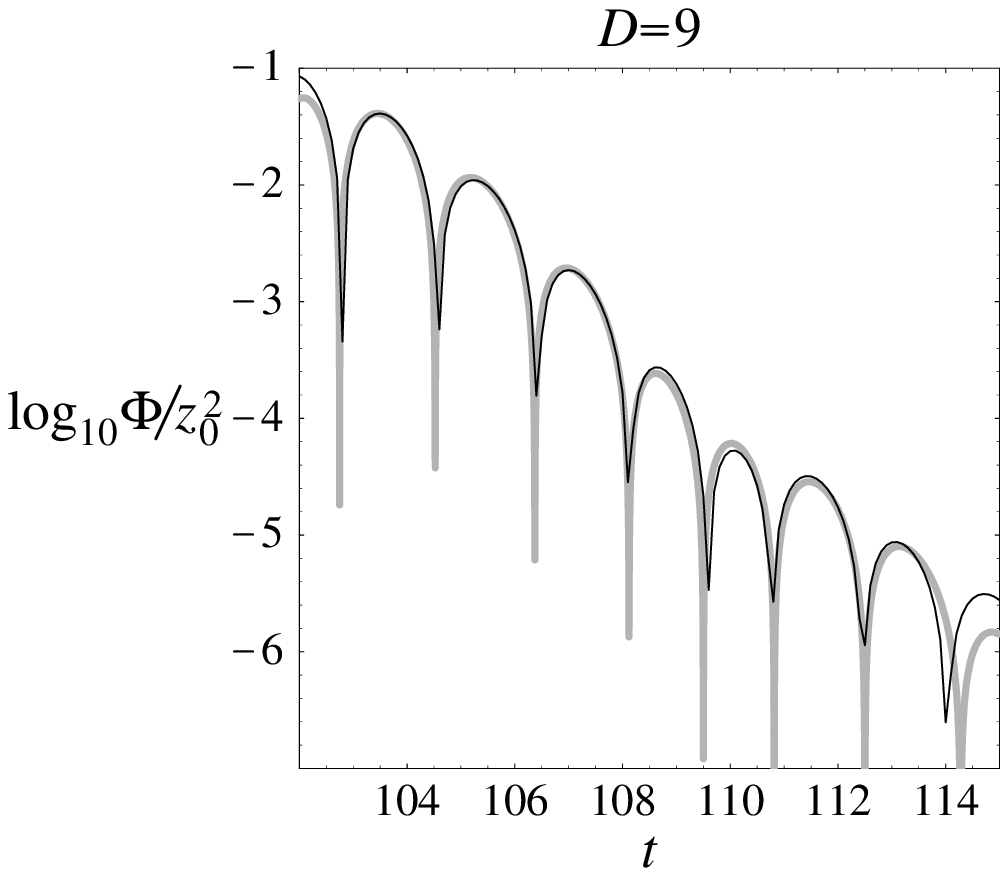}
\includegraphics[width=0.3\textwidth]{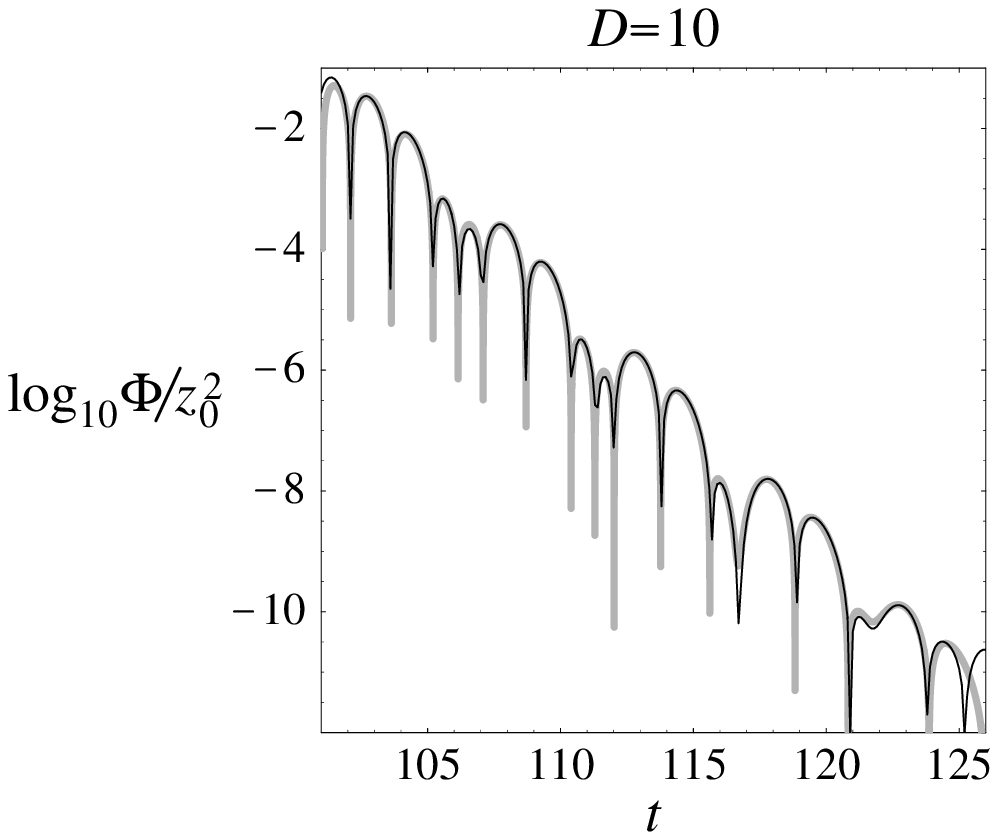}
\includegraphics[width=0.3\textwidth]{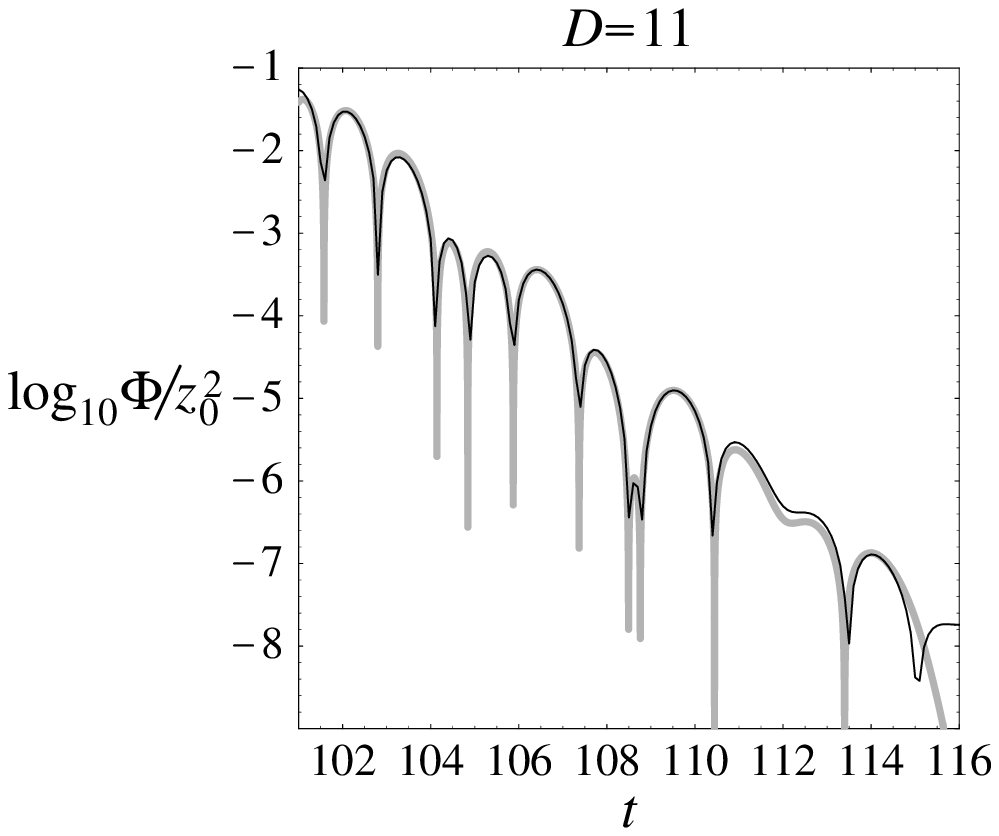}
}
\caption{The behavior of $\log_{10}\Phi(t)/z_0^2$ in the ringing phase
(the black line) and the fitted data with two values of $\omega_{\rm
QN}$ listed in Table~\ref{QNM-l2} (the gray line) for $D=9$, 10, and
$11$.}
\label{Fit}
\end{figure}

The values of $\omega_{\rm QN}$ are evaluated in the following manner:
The imaginary part ${\rm Im}(\omega_{\rm QN})$
is derived from the slope of the peaks of 
$\log \Phi(t)$ shown in Fig. 4. 
The real part is estimated by averaging the intervals of zeros of $\Phi(t)$
and consistency is checked by identifying the
Fourier peak of $\Phi(t)\times \exp(-{\rm Im}(\omega_{\rm QN})t)$.  
For $D=9,~10$, and $11$, we found that two modes are mixed 
and searched two values of $\omega_{\rm QN}$ 
so that the numerical data of $\Phi(t)$ is well fitted.
By comparing the results derived by using the several ranges of $t$,
we estimate the error to be $\lesssim 1\%$ for $4\le D\le 8$
and $\sim 5\%$ for $9\le D\le 11$.
We summarize the values of 
$\omega_{\rm QN}$ for $D=4$--11 in Table~\ref{QNM-l2}
and compare the fitted data and $\Phi(t)$ for $D=9$, 10, and 11 in
Fig.~\ref{Fit}.

The part of the derived results can be compared with previous
ones. For $D=4$, Leaver~\cite{L85}
gives very accurate values of
$\omega_{\rm QN}$ (see \cite{KS99} for a review).  
For the higher-dimensional Schwarzschild black holes, 
Konoplya~\cite{Konoplya} evaluated the values 
of $\omega_{\rm QN}$ for $l=2$ and $3$
using the WKB method and they were extended
to $l\ge 4$ in \cite{BCG04}. 
The Leaver's method
was applied to the scalar mode of the gravitational perturbation
for $D=5$ \cite{CLY03}. 
The values of $\omega_{\rm QN}$ derived
with these methods are summarized in Table~\ref{QNM-l2}. In the cases
$D=4$ and $5$, our results agree well with those in the previous studies. On
the other hand, for $D=6$, 8, and 10, our results disagree with the
previous ones.  As stated in \cite{Konoplya,BCG04}, the WKB method is not
expected to work well for $D=8$ and $10$ since the potential
$V_S$ has a negative peak.  Our results indicate that the WKB
method might not be good even for $D=6$.

\begin{table}[tb]
\centering
\caption{The same  as Table IV but for $l=4$. }
\begin{ruledtabular}
\begin{tabular}{c|ccc}
$D$ & Our estimate & Leaver's method & WKB  \\
  \hline 
4 & $1.618-0.188i$ & $1.6184-0.1883i$ & $1.618-0.188i$\\
5 & $2.193-0.328i$ & $2.1924-0.3293i$ & $\cdots$ \\
6 & $2.623-0.439i$ & $\cdots$ & $2.622-0.438i$\\
7 & $3.012-0.534i$ & $\cdots$ & $\cdots$ \\
8 & $3.389-0.631i$ & $\cdots$ & $3.401-0.645i$\\
9 & $3.779-0.734i$ & $\cdots$ & $\cdots$\\
10 & $4.176-0.838i$ & $\cdots$ & $4.223-0.841i$ \\
11 & $4.595-0.950i$ & $\cdots$ & $\cdots$\\
\end{tabular}
\end{ruledtabular}
\label{QNM-l4}
\end{table}

\begin{table}[tb]
\centering
\caption{The same  as Table IV but for $l=6$. }
\begin{ruledtabular}
\begin{tabular}{c|ccc}
$D$ & Our estimate & Leaver's method & WKB   \\
  \hline 
4 & $2.455-0.192i$ & $\cdots$ & $2.424-0.191i$\\
5 & $3.286-0.344i$ & $\cdots$ & $\cdots$ \\
6 & $3.913-0.470i$ & $\cdots$ & $3.911-0.467i$\\
7 & $4.610-0.574i$ & $\cdots$ & $\cdots$ \\
8 & $4.924-0.674i$ & $\cdots$ & $4.923-0.675i$\\
9 & $5.388-0.769i$ & $\cdots$ & $\cdots$\\
10 & $5.834-0.859i$ & $\cdots$ & $5.848-0.865i$ \\
11 & $6.292-0.955i$ & $\cdots$ & $\cdots$\\
\end{tabular}
\end{ruledtabular}
\label{QNM-l6}
\end{table}

Our methods can be applied to arbitrary values of $l$. To estimate
$\omega_{\rm QN}$, we evolved appropriate initial data for $l=4$ and
$6$. For these cases, only one quasinormal frequency appears in the
ring down phase and the error of $\omega_{\rm QN}$ is $\lesssim 1\%$
for all values of $D$.  Our results together with those in the
previous studies are summarized in Tables~\ref{QNM-l4} and
\ref{QNM-l6}.  They agree well within the difference $\lesssim 1\%$.
This implies that evolving appropriate initial data by the master
equation is an effective method for computation of the complex
frequencies of the fundamental quasinormal modes.


\section{Summary}

In this paper, we have studied gravitational waves emitted 
during head-on collision of two black holes in higher dimensions
using the close-limit approximation. 
We evolved the Brill-Lindquist initial data perturbatively, 
using a gauge-invariant techniques in a Schwarzschild black hole
and calculated the waveform and radiated energy $E_{\rm rad}$. 
$E_{\rm rad}$ is given by the formula~\eqref{radiated-E}
and the values of $\hat{E}_2$ are summarized in Table~\ref{Ehat}. 
At the critical separation for the presence of
the common apparent horizon $z_0=z_0^{\rm (crit)}$, our analysis
of the first order perturbation predicts that 
$E_{\rm rad}/M$ becomes larger with larger values of $D$.
There is a possibility that the higher-order correction is large for
$z_0=z_0^{\rm (crit)}$ 
in the higher-dimensional cases and to clarify the higher-order
effect is left as a remaining problem. 
We also evaluated the values of $\alpha_D=E_{\rm rad}/(M-M_{\rm AH})$
at $z_0\ll 1$ and found that $\alpha_D$ ($5\le D\le 11$) is larger 
than $\alpha_4$. These results indicate that more energy could
be radiated away in higher dimensional spacetimes than  in the
four dimensional one during head-on collision with the
approaching velocity much smaller than the speed of light.

It has also been illustrated that the fundamental quasinormal mode
frequencies for the scalar-mode perturbation in the Schwarzschild
black holes can be computed in our analysis. We derived the complex
frequencies for various values of $D$, which have not been accurately
computed so far.

As we mentioned in the introduction, the close-limit analysis for
head-on collision of the black holes performed in this paper is the
first step toward more general studies of the black hole collision in
higher dimensions.  As the next step, we plan to analyze the evolution
of two initially-moving black holes using the close-slow
approximation. Both the head-on collisions and the grazing collisions
should be studied. As the momentum of each black hole increases,
dependence of the emissivity of gravitational waves on the value of $D$
would be changed from the results in this
paper. By observing such behaviors, we will be able to discuss the
dependence of gravitational radiation in the collision
process on the value of $D$ in a different way from the previous studies.

\acknowledgments

This work was supported by Grant-in-Aid for Scientific 
Research from Ministry of Education, Science, Sports, and Culture of 
Japan (Nos. 13135208, 14740155, 14102004, 17740136, and 17340075). 

\appendix

\section{Determining the initial master variable}

In this section, we explain how to match the initial master variable
$\Phi$ to the initial data. We begin by briefly reviewing the gauge
invariant formulation of the perturbation in the Schwarzschild black hole
spacetime~\cite{KI03}.  The background metric is given by
\begin{equation}
ds^2=g_{ab}dy^ady^b+r^2(y)\gamma_{ij}dz^idz^j,
\end{equation}
where $g_{ab}dy^ady^b=-fdt^2+f^{-1}dr^2$ and $\gamma_{ij}dz^idz^j$
denotes the metric on a unit sphere. The perturbed metric 
is written as
\begin{equation}
ds^2=(g_{ab}+h_{ab})dy^ady^b+
(h_{ai}+h_{ia})dy^adz^i
+(r^2\gamma_{ij}+h_{ij})dz^idz^j.
\end{equation}
The perturbation variables $h_{ab}$, $h_{ai}$, and $h_{ij}$
can be separated into scalar, vector, and tensor modes.
Each mode is further expanded into the modes of
different anuglar quantum number $l=0$, $1, \cdots$ using the
hyper-spherical harmonics ${\mathbb S}$, 
which is the solution of the following equation:
\begin{equation}
(\hat{D}_i\hat{D}^i+k^2){\mathbb S}=0.
\end{equation}
Here $\hat{D}^i$ denotes the covariant derivative on the unit sphere
and the definition of $k^2$ is given in Eq.~\eqref{kk-and-x}.
The variables of the scalar mode perturbation are given by
\begin{align}
h_{ab}&=f_{ab}{\mathbb S},\\
h_{ai}&=rf_a{\mathbb S}_i,\\
h_{ij}&=2r^2(H_L\gamma_{ij}{\mathbb S}+H_T{\mathbb S}_{ij}),
\end{align}
where
\begin{align}
&{\mathbb S}_i=-\frac{1}{k}\hat{D}_i{\mathbb S},\\
&{\mathbb S}_{ij}=\frac{1}{k^2}\hat{D}_i\hat{D}_j{\mathbb S}
+\frac{1}{n}\gamma_{ij}{\mathbb S}.
\end{align}
In the axisymmetric case the metric of a unit sphere is written as 
\begin{equation}
\gamma_{ij}dz^idz^j=d\theta^2+\sin^2\theta d\Omega_{n-1}^2,
\end{equation}
and then 
\begin{equation}
{\mathbb S}=S^{[n]}_l
\equiv K^{[n]}_lC^{[(n-1)/2]}_l(\cos\theta),
\end{equation}
\begin{equation}
K^{[n]}_{l}=\left[\frac{4\pi^{(n+1)/2}\Gamma(n+l-1)}
{(n+2l-1)\Gamma(l+1)\Gamma((n-1)/2)\Gamma(n-1)}\right]^{-1/2}.
\end{equation}
$K^{[n]}_{l}$ is the normalization factor that is chosen so that
\begin{equation}
\int S^{[n]}_lS^{[n]}_{l^\prime}d\Omega_n=\delta_{ll^\prime}
\end{equation}
is satisfied. For the gauge transformation generated
by the following vector fields,
\begin{equation}
\xi_a=T_a{\mathbb S},~~~\xi_i=rL{\mathbb S}_i,
\end{equation}
the gauge invariant quantities of the perturbation
are given by
\begin{align}
&F=H_L+(1/n)H_T+(1/r)D^arX_a,\\
&F_{ab}=f_{ab}+D_aX_b+D_bX_a,
\end{align}
where
\begin{equation}
X_a=\frac{r}{k}\left(f_a+\frac{r}{k}D_aH_T\right).
\end{equation}
The master variable $\Phi$ is related to the gauge invariant quantities as follows:
\begin{eqnarray}
X&\equiv r^{n-2}\left({F}^t_t-2{F}\right)
&=r^{n/2-2}\left(-\frac{r^2}{f}\partial_t^2\Phi
-\frac{P_X}{16H^2}\Phi+\frac{Q_X}{4H}r\partial_r\Phi\right),\\
Y&\equiv r^{n-2}\left({F}^r_r-2{F}\right)
&=r^{n/2-2}\left(\frac{r^2}{f}\partial_t^2\Phi
-\frac{P_Y}{16H^2}\Phi+\frac{Q_Y}{4H}r\partial_r\Phi\right),\\
Z&\equiv r^{n-2}{F}^r_t
&=r^{n/2-1}\left(-\frac{P_Z}{4H}\partial_t\Phi
+fr\partial_r\partial_t\Phi\right),
\end{eqnarray}
where
\begin{eqnarray}
&P_X(r)=
 & n^3(n+1)^3x^3+2n(n+1)[2(n^2+n+2)m-n(n-2)(n+1)]x^2
\notag\\
&& -4n[(n-11)m+n(n+1)(n-3) ]mx+16m^3+8m^2n^2,\label{PX}\\
&Q_X(r)=
 & n(n+1)^2x^2+2[(3n-1)m-n(n+1)]x-4nm,\label{QX}\\
&P_Y(r)=
& n^3(n-1)(n+1)^2x^3+2n(n^2-1)[4m-n(n-2)(n+1)]x^2 \notag\\
&& +4n(n-1)[3m+n(n+1)]mx,\label{PY}\\
&Q_Y(r)= 
 & n(n-1)(n+1)x^2 -2(n-1)[m+n(n+1)]x,\label{QY}\\
&P_Z(r)=
 & [-n^2(n+1)x+2(n-2)m]y+n(n+1)x^2 \notag\\
&& +[2(2n-1)m+n(n+1)(n-2)]x -2nm.
\end{eqnarray}
Here, the definition of $x$ is given in Eq.~\eqref{kk-and-x}.

Now we turn our attention to the method for determining the initial
master variable $\Phi$ from a full nonlinear time symmetric initial
data. We denote $f_{ab}$ and $f_a$ as
\begin{equation}
f_{ab}=\left(
\begin{array}{cc}
fH_0&H_1\\
H_1&f^{-1}H_2
\end{array}
\right),~~~
rf_a=(h_0,h_1).
\label{fab-fa}
\end{equation}
Because of the time symmetry of the initial condition, 
\begin{equation}
\dot{H}_0=\dot{H}_2=\dot{h}_1=\dot{H}_L=\dot{H}_T=0,
\end{equation}
\begin{equation}
H_1=h_0=0.
\end{equation}
By the comparison with Eqs.~\eqref{BL_Schwarzschild} and
\eqref{fac_BL}, we find
\begin{equation}
H_2=2H_L=\chi(r)\equiv\frac{1/(n-1)R^{n-1}}{1+1/4R^{n-1}}
\left(\frac{z_0}{R}\right)^l\left(K_l^{[n]}\right)^{-1},
\label{chi}
\end{equation}
\begin{equation}
h_1=H_T=0. 
\end{equation}
Then the gauge-invariant quantity is found to be 
\begin{equation}
F=H_L,~~~F^t_r=0,
\end{equation} 
and thus
\begin{equation}
X+Y=-nr^{n-2}\chi(r),
\label{XplusY_chi}
\end{equation}
where we have used one of Einstein's equation
$F^a_a=-2(n-2)F$. On the other hand, $X+Y$ is given in terms of $\Phi$
as follows:
\begin{equation}
X+Y=r^{n/2-2}\left(
-\frac{P_X+P_Y}{16H^2}\Phi
+\frac{Q_X+Q_Y}{4H}r\partial_r\Phi
\right).
\label{XplusY}
\end{equation}
Hence we find the following equation for the initial master variable $\Phi$
\begin{equation}
\frac{d\Phi}{dr_*}=\frac{f}{r(Q_X+Q_Y)}
\left[
\frac{P_X+P_Y}{4H}\Phi-4nr^{n/2}H\chi(r)
\right].
\label{initial_eq}
\end{equation}
We also find $\dot{\Phi}=0$ from the condition $Z=0$.

Taking the limit $r_*\to-\infty$ of Eq.~\eqref{initial_eq}, we find
\begin{equation}
\Phi=\frac{2\cdot 4^{l/(n-1)}nz_0^l}{(n-1)(n+m)K_l^{[n]}}. 
\label{initial_phi_asymptotic}
\end{equation}
This gives us the boundary condition at $r_* = -\infty$. 
We solve Eq. \eqref{initial_eq} using the fourth-order
Runge-Kutta method with the boundary condition (A34).

\section{Formula of radiated energy}

In this section, we sketch the derivation of
Eq. \eqref{radiated_energy_master_variable}. The similar calculation
for the energy spectrum is found in Ref. \cite{BCG04}.

In the region far from the source, outgoing waves are
well approximated by the spherical ones in the transverse-traceless
(TT). Then, the perturbation is written as 
\begin{equation}
h_{ij}^{\rm TT}\simeq 2r^2H_T{\mathbb S}_{ij},
\end{equation}
\begin{equation}
H_T\simeq \frac{A}{r^{n/2}}h(t-r).
\label{asymptotic_behavior}
\end{equation}
The radiated energy flux is given by
\begin{equation}
\frac{dE}{dSdt}=\frac{1}{32\pi G}{\dot{h}^{\rm TT}_{ij}}\dot{h}^{{\rm TT} ij}
=\frac{1}{8\pi G}\dot{H}_T^2
{\mathbb S}_{ij}{\mathbb S}^{ij},
\label{radiation1}
\end{equation}
which is the same form as in the four-dimensional case.
Using the formula \cite{BCG04}
\begin{equation}
{\mathbb S}_{ij}{\mathbb S}^{ij}=
\frac{1}{k^4}
\hat{D}_i[\hat{D}_j{\mathbb S}\hat{D}^i\hat{D}^j{\mathbb S}
+(k^2-n+1){\mathbb S}D^i{\mathbb S}]
+
\frac{(k^2-n)(n-1)}{k^2n}{\mathbb S}^2,
\label{SijSij}
\end{equation}
we find
\begin{equation}
\frac{dE}{dt}
=r^n\frac{\dot{H}_T^2}{8\pi G}
\frac{(n-1)(k^2-n)}{nk^2}.
\label{radiation2}
\end{equation}

Next we rewrite the formula \eqref{radiation2} in terms of the
master variable $\Phi$. In a region far from the source, the
gauge invariant quantities become 
\begin{equation}
F=\frac{1}{n}H_T+\frac{f}{r}X_r,~~~F_{ab}=D_aX_b+D_bX_a,
\end{equation}
\begin{equation}
X_a=\frac{r^2}{k^2}\partial_a H_T.
\end{equation}
If we calculate $F$, $F^t_t$, $F^r_r$, and $F^r_t$ keeping only the
leading order $O(r^{2-n/2})$ and the subleading order $O(r^{1-n/2})$,
we find
\begin{equation}
Y+Z=n\frac{r^{n-1}}{k^2}\dot{H}_T,
\end{equation}
where Eq.~\eqref{asymptotic_behavior} was also used.
On the other hand, calculating $Y+Z$ in terms of $\Phi$ and using
the fact that $\ddot{\Phi}=-\partial_r\dot{\Phi}$ holds for the outgoing wave,
we obtain 
\begin{equation}
\dot{H}_T=\frac{k^2}{2}r^{-n/2}\dot{\Phi}.
\label{HT-Phi}
\end{equation}
Substituting this equation into Eq.~\eqref{radiation2}, we find
\begin{equation}
E_{\rm rad}=\sum_l\frac{k^2(n-1)(k^2-n)}{32\pi nG}\int\dot{\Phi}^2dt.
\label{Erad-formula}
\end{equation}
This formula is equivalent to Eq.~\eqref{radiated_energy_master_variable}
in the unit $r_h(M)=1$.

Using Eqs.~\eqref{radiation1}, \eqref{HT-Phi} and \eqref{Erad-formula},
we find
\begin{equation}
\frac{1}{E}\frac{dE}{d\Omega_n}=\frac{1}{k^2(k^2-n)(n-1)^2}
\left(n{\mathbb S}_{,\theta\theta}+k^2{\mathbb S}\right)^2.
\end{equation} 
In the case $l=2$ it becomes
\begin{equation}
\frac{1}{E}\frac{dE}{d\Omega_n}=
\frac{2\pi^{-(n+1)/2}\Gamma((n+5)/2)}{n(n+2)}\sin^4\theta,
\end{equation}
which reduces to Eq.~\eqref{angular-dependence} using 
$d\Omega_n=\Omega_{n-1}\sin^{n-1}\theta d\theta$.

\end{document}